upon request 

\documentstyle[aps,preprint]{revtex}
\begin{document}

\author{Alexander G. Abanov}
\address{James Franck Institute, the University of Chicago,
         5640 S. Ellis Ave, Chicago, IL 60637, U.S.A.}
\author{Oleg A. Petrenko}
\address{Department of Physics and Astronomy, McMaster University,
 1280 Main Str. West, Hamilton, Ontario, Canada L8S 4M1}

\preprint{cond-mat/9402024}

\title{Enhancement of Anisotropy Due to Fluctuations in
Quasi-One-Dimensional Antiferromagnets}

\newcommand{\rmb}{RbMnBr$_3$}
\newcommand{\cnc}{CsNiCl$_3$}
\newcommand{\knc}{KNiCl$_3$}
\newcommand{\cmb}{CsMnBr$_3$}
\newcommand{\abx}{{\bf ABX$_3$}}
\newcommand{\rnc}{RbNiCl$_3$}

\maketitle

\begin{abstract}
It is shown that the observed anisotropy of magnetization at high
 magnetic fields in \rmb , a quasi-one-dimensional antiferromagnet
on a distorted stacked triangular lattice, is due to quantum and thermal
fluctuations. These fluctuations are taken into account in the framework of
linear spin-wave theory in the region of strong magnetic fields. In this
region the divergent one-dimensional integrals are cut off by magnetic
field and the bare easy-plane anisotropy. Logarithmical dependence on the
cutoff leads to the "enhancement" of the anisotropy in magnetization.
Comparison between magnetization data and our theory with parameters
obtained from neutron scattering experiments has been done.
\end{abstract}

\section{Introduction}
The question of the effect of fluctuations on the magnetic properties of
quasi-one-dimensional antiferromagnets has been intensively discussed in the
last few years. Particular attention has been given to the cases where
there is a stacked triangular lattice and an antiferromagnetic interaction, so
that there is a frustration in going from one-dimensional to three-dimensional
ordering.

This situation is apparent in
materials with hexagonal \cnc -type crystal  structure having the general
formula \abx\, where {\bf A} is an alkali metal, {\bf B} is a bivalent metal of
3d group, and {\bf X} is a halogen. While spin dynamics  of \cnc\ and \rnc\
(S=1, easy-axis anisotropy) has been studied to examine Haldane  conjecture
\cite{Haldane,Morra}, the compounds of the same group, \cmb ,
\rmb\ (S=5/2, easy-plane anisotropy) can be considered as appropriate for
checking the applicability of standard spin-wave theory for the case of large
half-integer spins.

\cmb\ has been investigated by various experimental techniques as an example
of a frustrated antiferromagnet on a stacked triangular lattice exhibiting a
number  of unusual magnetic properties, such as a field-induced phase
transition
from the triangular phase to the collinear one, critical behavior
associated with
a chiral degeneracy and an unusual phase diagram. In addition to numerous
neutron  scattering experiments \cite{neutrCollins} and to ESR measurements
\cite{cmbresonans}, the magnetization process has been studied in detail
\cite{Chubukov,Beba,Jap,AbarBazh} with results which are in good agreement
with classical calculations \cite{Chubukov} except for two discrepancies.
Namely:
\begin{itemize}
\item the measured magnetic torques are significantly smaller
than the theoretical ones; \item in large magnetic fields ($H>H_c$, where
$H_c$ is the critical field of the transition to the collinear phase) there
is a considerable anisotropy between magnetization when magnetic field is
applied along C-axis of the crystal ($M^{H\parallel C}$) and the one
($M^{H\perp
C}$) for magnetic field along easy plane absent in the theory \cite{Chubukov}.
\end{itemize}

It was suggested \cite{AbarBazh} that in order to describe these
peculiarities the anisotropy of quantum fluctuations should be taken into
account. In this paper we consider on the basis of another compound \rmb\ how
both quantum and thermal fluctuations affect the magnetization of
quasi-one-dimensional noncollinear antiferromagnet.

\section{Crystal structure and magnetic pro\-per\-ties of \rmb }

Powder neutron scattering measurements by Glinka et al. \cite{Glinka}
showed that \rmb\ exhibits antiferromagnetic order below T$_N=8.8\pm0.1$K.
Single crystal measurements \cite{Collins,Kato_neutr} show that the magnetic
structure of \rmb\ is incommensurate with $Mn^{+2}$ moments lying in the basal
plane and with antiferromagnetic ordering in the $c$ direction. The
commensurate
ordering in \cmb\ gives scattering peaks of type ($\frac{1}{3}$, $\frac{1}{3}$,
$1$) \cite{neutrCollins} while in the incommensurate \rmb\ these peaks are
replaced by two triads of peaks near to ($\frac{1}{3}$, $\frac{1}{3}$, $1$)
with
three fold symmetry around ($\frac{1}{3}$, $\frac{1}{3}$, $1$). Because the
incommensurate peaks surround a commensurate position in reciprocal space, it
is
likely that the incommensurate structure is locally similar to the
120$^{\circ}$
triangular structure of \cmb , but with angles between neighboring spins in the
basal plane somewhat larger than 120$^{\circ}$.

The appearance of incommensurate magnetic structure may be due to the presence
 of crystal structure distortions. According to X-ray study by Von Fink and
Seifert \cite{Fink} as the temperature decreases the crystal structure of \rmb\
changes at 470K from \cnc -type structure (space group $P6_3/mmc$) to the
so-called \knc -type structure (space group $P6_3cm$). Measurements of the
birefringence \cite{Kato} and recent neutron scattering experiments \cite{YMT}
show that another structural phase  transition takes place at about 220K. As
reported in \cite{YMT} below this temperature the crystal has orthorhombic
structure with cell dimensions $a$=14.680\AA, $b$=12.805\AA, $c$=6.516\AA\ at
12K.

At low temperatures a field of 3.0T applied along the $a$ axis produces a
first-order phase transition to a commensurate structure
\cite{Kato_neutr,Kawano}, so that the lattice parameter of the magnetic cell is
eight times that of the nuclear cell \cite{Collins}. In a magnetic field of
about $H_c=3.9$T and above applied  in the basal plane, the magnetic structure
becomes collinear, as in the case of \cmb\ with $H_c=6.4$T. This conclusion can
be derived from a previous study of the magnetization process \cite{my-stat}
and
ESR-spectrum \cite{my-res} in \rmb .

The crystal distortions lead primarily to a change in the antiferromagnet
coupling between chains. But for the purpose of this paper the  small
distinction
between the \rmb\ magnetic structure and a simple 120$^{\circ}$-structure
is not so important. Moreover confining ourselves by
the range of large magnetic fields we assume that the
effects of inter-chain exchanges and particularly the distortion of a simple
120$^{\circ}$ magnetic structure in the absence of magnetic fields will give
only quantitative corrections not changing the qualitative picture. This
assumption appears to be true and we estimate these corrections for large
magnetic fields (see (\ref{final}-\ref{3perp}) below).

\section{Experimental procedure and results}

A vibration-sample magnetometer, similar to the one described in
ref.\cite{Bazhan}, was used to measure the magnetization in \rmb . A magnetic
field up to 6T was generated by two superconducting coils. Simultaneous
measurements of two mutually perpendicular components of the magnetization of
the sample, one of which ($M_{long}$) is parallel and another ($M_{tr}$)
is perpendicular to the direction of magnetic field were done by using three
pairs of measuring coils.

The absolute accuracy of  the magnetization measurements was  about 5\%.
The crystal was oriented with an accuracy of 1-2$^{\circ}$.

The investigation was performed on single crystals having approximately $1.5
\times 1.5 \times 1.5 mm^3$ volume and approximately 30mg mass in the
temperature range 1.7-12K.

The magnetization  component $M_{long}$ parallel to the field as a function of
the field $H$ for $H \!\! \parallel \!\! C$ and $H \!\! \perp \!\! C$ at
T=1.7K is
shown in Fig.1. In fields above the critical field ($H_c\approx  3.9$T) the
magnetizations $M_{long}^{H\parallel C}$ and $M_{long}^{H\perp C}$ are not the
same, but differ by about 7\%. This anisotropy of magnetization can not be
explained just by taking into account the easy plane anisotropy in classical
calculations, because it leads to $M_{long}^{H\parallel C} <
M_{long}^{H\perp C}$
while experimentally the situation is opposite. Moreover corrections to the
classical magnetization due to easy-plane anisotropy have to be of order $D/J
\approx 1\%$ which is much smaller than the observed anisotropy ($5-10\%$).

The anisotropy in magnetization is shown on Fig.2 where the magnetization
component $M_{tr}$ perpendicular to the magnetic field $H$ as a function of the
field is plotted at different angles $\varphi$ between the field and the basal
plane and at T=4.2K. Due to the  anisotropy of the magnetization at $H>H_c$,
a non zero transverse-magnetization signal occurs at $\varphi \neq 0$.

According to Fig.2 the field dependence $M_{tr}(H)$ exhibits a small
hysteresis.
This hysteresis takes place when the projection of the magnetic field on the
basal plane reaches the value of $3.19 \pm 0.10$T in increasing fields and
$2.67 \pm 0.07$T in decreasing fields. It corresponds to the first order phase
transition from the incommensurate magnetic structure to the commensurate one.

In Fig.3 the field dependence of the magnetization component $M_{tr}$ at
different
temperatures is shown. At large magnetic fields we can see that $M_{tr}$
increases
with temperature supporting the idea that the anisotropy of the magnetization
at
$H>H_c$ is due to fluctuations.

This is even clearer from Fig.4 where the value of $M_{tr}(H=4.5$T,
$\varphi = 20^{\circ})$ is shown as a function of temperature.

\section{Theory}

We start with the microscopic Hamiltonian
\begin{eqnarray}
 \cal{H}  &  =   &   2J \sum_i \vec{S}_i\vec{S}_{i+\Delta_z} +
   2J' \sum_i \vec{S}_i\vec{S}_{i+\Delta_{\perp}}
    + D \sum_i (S_{i}^{z})^2 - \vec{h}\sum_i \vec{S_i}
\label{modHam}
\end{eqnarray}
The intra-chain antiferromagnet exchange constant $J$ is the largest
parameter in
the Hamiltonian~(\ref{modHam}) and is about two orders of magnitude larger than
the exchange constant between chains $J'$ and easy-plane single-ion anisotropy
$D$ \cite{Collins}, $J>>J',D$ and $J,J',D>0$. $\vec{h}$ is the magnetic
field in
the units of energy and it is equal to
$ h=g \mu_B H$ where $H$ is a magnetic field in some standard units, $\mu_B$ is
a Bohr magneton and $g$ is Lande factor which is equal 2 in the compound under
consideration. For $h=0$ the classical ground state of the
Hamiltonian~(\ref{modHam}) is the $120^{\circ}$ structure with spin vectors
forming equilateral triangles in the basal plane. The case of a non zero
magnetic
field was considered classically by Chubukov \cite{Chubukov}. For $\vec{h}$
along
the Z ($C$) axis the transverse components of the  spins conserve $120^{\circ}$
structure and the classical magnetization
\begin{equation}
M_{h \parallel C} \equiv <S^z>=\frac{h}{8J}
\end{equation}
when $h < 8JS$ neglecting by $J'/J, D/J$ (here and
there on $M$ means $M_{long}$ where the opposite is not stated explicitly).
The case of the magnetic field perpendicular to the Z axis is more complicated.
For a sufficiently strong easy-plane anisotropy, spins do not leave the
planes but
the $120^{\circ}$ structure is not conserved anymore and a spin flip of the two
magnetic sublattices takes place at the relatively small magnetic field $h_c =
\sqrt{48JJ'S^2}$. At $h_c<h<8JS$ we have a collinear spin structure and
magnetization has to be the same (up to small $\frac{J'}{J}, \frac{D}{J}$
corrections) as in the case of field along Z axis \cite{Chubukov}.

In this section we take into account quantum and thermal fluctuations in the
framework of linear spin-wave theory and show that the anisotropy of
magnetization at $h>h_c$ due to these fluctuations is of the right sign
($M_{h\perp C}<M_{h\parallel C}$) and is much larger than the one expected from
classical calculation.

In what follows we confine ourselves to magnetic fields $h>h_c$ where we have a
collinear spin  structure for a magnetic field perpendicular to the $C$ axis of
the crystal and we neglect the $J'$ term in the Hamiltonian~(\ref{modHam})
in the
body of this chapter. This term will give corrections to the gap in a spin-wave
spectrum of order of $J'/J$ without  changing the qualitative picture (see
(\ref{3par}-\ref{3perp}) below). Thus  we consider the Hamiltonian of a single
magnetic chain instead of (\ref{modHam}) keeping in mind that our results
will be
applicable only for the $h>h_c = \sqrt{48JJ'S^2}$ region. We can write
Hamiltonians for two directions of magnetic field in the form:
\begin{eqnarray}
 {\cal H}_{h \parallel C}  &  =   &   2J \sum_i \vec{S}_i\vec{S}_{i+1}
          - h\sum_i S_i^z + D \sum_i (S_{i}^{z})^2
\label{parHam}  \\
 {\cal H}_{h \perp C}      &  =   &   2J \sum_i \vec{S}_i\vec{S}_{i+1}
          - h\sum_i S_i^z + D \sum_i (S_{i}^{x})^2
\label{perHam}
\end{eqnarray}
Here we slightly changed notations for simplicity. Now magnetic field is
always applied along Z axis and the easy plane is XY plane for the parallel
case and YZ plane for the perpendicular one. Let us apply usual Dyson-Maleev
transformation
\begin{eqnarray}
 S^{z'}  &  =   &   -S + a^{+}a    \nonumber \\
 S^+  &  =   &   \sqrt{2S}a^+(1-\frac{a^+a}{2S}) \\
 S^-  &  =   &   \sqrt{2S}a  \nonumber
\end{eqnarray}
to (\ref{parHam},\ref{perHam}) with the axes of quantization $z'$ directed
as it
is shown in Fig. 5. Choosing the angle $\phi$ to cancel linear in $a, a^+$
terms
in Hamiltonians~(\ref{parHam},\ref{perHam}) we reproduce the classical
results for
magnetization. The terms of (\ref{parHam}, \ref{perHam}) quadratic in $a, a^+$
will give us the linear spin-wave theory corrections to these results. We have:
\begin{eqnarray}
 \sin\phi_{\parallel} &  =  &-\frac{h}{2S(4J+D)} \label{phipar}
  \\
 \sin\phi_{\perp}     &  =  &-\frac{h}{8JS}      \label{phiperp}
\end{eqnarray}
and both Hamiltonians (\ref{parHam}, \ref{perHam}) after Fourie transformation
can be written in the form
\begin{eqnarray}
 \cal{H} &  =  &\gamma + \sum_k \left[ \omega_{1}(k)a_k^+a_k-
           \frac{\omega_{2}(k)}{2}(a_k^+a_{-k}^++a_ka_{-k})\right]
\label{formHam}
\end{eqnarray}
with
\begin{eqnarray}
 \gamma_{\parallel}      & = &-N(2JS^2(1+2\sin^2\phi)-\frac{DS}{2}\cos^2\phi
                              +DS^2\sin^2\phi)         \nonumber\\
 \omega_{1\parallel}(k) & = & 4JS(1-\sin^2\phi\cos k)+DS\cos^2\phi
 \nonumber\\
 \omega_{2\parallel}(k) & = & 4JS\cos^2\phi\cos k+DS\cos^2\phi
     \nonumber\\
 \gamma_{\perp}          & = & -N(2JS^2(1+2\sin^2\phi)-\frac{DS}{2}
  \nonumber\\
 \omega_{1\perp}(k)     & = & 4JS(1-\sin^2\phi\cos k)+DS
     \nonumber\\
 \omega_{2\perp}(k)     & = & 4JS\cos^2\phi\cos k-DS
     \nonumber
\end{eqnarray}
Making Bogolyubov transformation
\begin{eqnarray}
a_k & = & \left(\frac{\omega_1+\sqrt{\omega_1^2-\omega_2^2}}
{2\sqrt{\omega_1^2-\omega_2^2}}\right)^{1/2}b_k+
\left(\frac{\omega_1-\sqrt{\omega_1^2-\omega_2^2}}
{2\sqrt{\omega_1^2-\omega_2^2}}\right)^{1/2}b_{-k}^+
\end{eqnarray}
we diagonalize (\ref{formHam}) and get
\begin{equation}
{\cal H} =
\gamma + \sum_k \left[\epsilon(k)(b_k^+b_k+\frac{1}{2})
           -\frac{\omega_1(k)}{2} \right]
\end{equation}
with the spectrum of spin waves given by
\begin{displaymath}
 \epsilon^2(k) = \omega_1^2(k)-\omega_2^2(k)
\end{displaymath}
or
\begin{eqnarray}
 \epsilon^2_{\parallel}(k) & = & 4JS(1-\cos k)
                                 \left[4JS(1+\cos2\phi\cos k)
                                 +2DS\cos^2\phi\right]
                                 \label{epspar} \\
 \epsilon^2_{\perp}(k)     & = & \left[4JS(1-\cos k)+2DS\right]
                                 4JS(1+\cos2\phi\cos k)
                                 \label{epsperp}
\end{eqnarray}
Now we can find the magnetization using general relation
$M=-<\frac{\partial{\cal H}}{\partial h}>$
\begin{equation}
M = -\frac{\partial \gamma}{\partial h} + \sum_k
\left[-\frac{\partial \epsilon(k)}{\partial h}(<b_k^+b_k>+\frac{1}{2})
           +\frac{1}{2}\frac{\partial \omega_1}{\partial h} \right]
\end{equation}
or after simple manipulations using
$(<b_k^+b_k>+\frac{1}{2})=\coth\frac{\epsilon}{2T}$
\begin{equation}
M=-S\sin\phi -
\int^{\pi}_{-\pi}\frac{dk}{2\pi}\left[-\frac{\partial \epsilon(k)}{\partial
 h}
\right]\frac{1}{2}\coth\frac{\epsilon}{2T}
\label{genmagn}
\end{equation}
Here the first term describes the classical part of magnetization while the
second one comes from the contribution of quantum and thermal fluctuations.

Let us show now that Eq.\ (\ref{genmagn}) leads to the enhanced anisotropy
between $M_{h \parallel C}$ and $M_{h \perp C}$. Consider the case of zero
temperature for simplicity. Substituting $1$ for $\coth\frac{\epsilon}{2T}$ and
taking derivatives using
(\ref{phipar},\ref{phiperp},\ref{epspar},\ref{epsperp})
we have  \begin{eqnarray}
 M_{h \parallel C} &  =  &-S\sin\phi \nonumber \\
               &     &-\sin\phi \int_{-\pi}^{\pi}\frac{dk}{2\pi}
                     \frac{1}{2}\sqrt{\frac{(1-\cos k)}
                      {1+\cos2\phi\cos k+D/2J\cos^2\phi}}
                      \frac{\cos k+D/4J}{1+D/4J}     \nonumber \\
 M_{h \perp C}     &  =  &-S\sin\phi-\sin\phi
                    \int_{-\pi}^{\pi}\frac{dk}{2\pi}
                     \frac{1}{2}\sqrt{\frac{1-\cos k+D/2J}
                     {1+\cos2\phi\cos k}}\cos k
\label{magn}
\end{eqnarray}
Now it is clear that only one of two soft modes of spin waves corresponding
$k\!=\!\pi$ gives a big contribution to the magnetization. This result is quite
obvious because it is this mode that corresponds to fluctuations of the angle
between spins and magnetic field. Another soft mode at $k\!=\!0$ corresponds to
asimuthal fluctuations of spins around the direction of magnetic field and
almost does not contribute to magnetization. This mode however contributes
very much to the value of average spin on the site.\footnote{It is seen
from this
argument that the attempts \cite{Zaliznyak} to relate the suppression of
magnetization with the suppression of average spin by fluctuations are
qualitatively wrong because these effects arise from the different modes of
spin
waves.} Expanding the integrands in  Eq.(\ref{magn}) in the vicinity of
$k\!=\!\pi$
and neglecting $D/J$ in comparison with $1$ we get  \begin{equation}
 M \approx -S\sin\phi +\sin\phi
\int\frac{dk}{2\pi}\frac{1}{\sqrt{k^2+g^2}}
\label{integral}
\end{equation}
with
\begin{eqnarray}
g_{\parallel}^2 & = & \frac{h^2}{16J^2S^2}+\frac{D}{J}  \\
g_{\perp}^2     & = & \frac{h^2}{16J^2S^2}
\end{eqnarray}
or taking integrals finally
\begin{eqnarray}
M_{h \parallel C} & \approx & \frac{h}{8J}\left(1+\frac{1}{2\pi S}
                          \ln(\frac{h^2}{16J^2S^2}+\frac{D}{J})\right)
\label{final}\\
M_{h \perp C}     & \approx & \frac{h}{8J}\left(1+\frac{1}{2\pi S}
                          \ln(\frac{h^2}{16J^2S^2})\right)
\label{final1}
\end{eqnarray}
Now let us estimate the corrections to (\ref{final}-\ref{final1}) due to 3D
($J'$)
effects. These effects give corrections of order of $J'/J$ under logarithm in
(\ref{final}-\ref{final1}) \footnote{Expressions like $\ln{\frac{J'}{J}}$ were
obtained in \cite{Welz} for spin reduction but can not be applied to our
case (see
footnote 1).} and can be calculated  in the framework of the same linear
spin wave
theory. Calculations show \cite{Abanov} that 3D fluctuations will
effectively increase the  gap of 1D system giving instead of
(\ref{final}-\ref{final1}) :
\begin{eqnarray} M_{h\parallel C} & = &
\frac{h}{8J}\left[ 1+\frac{1}{2\pi S}\ln \left(
\frac{h^2}{16J^2S^2}+\frac{D}{J}+3\frac{J'}{J}\right)\right] \label{3par}\\
M_{h\perp C} & = & \frac{h}{8J}\left[ 1+\frac{1}{2\pi S}\ln \left(
\frac{h^2}{16J^2S^2}+2\frac{J'}{J}\right)\right] \label{3perp}
\end{eqnarray} The
formulas (\ref{3par}-\ref{3perp}) are good at sufficiently big magnetic fields
where fluctuations and effects of three-dimensional ordering are small enough
to
guarantee the applicability of linear spin wave theory for
quasi-one-dimensional
system.

Now from (\ref{final}-\ref{3perp}) the origin of the enhancement of
anisotropy is clear. Big fluctuations of the  angle between spins and magnetic
field give a significant contribution to magnetization which depends strongly
on
the corresponding gap in the spectrum of spin waves due to low dimensionality
of
the system (integral in Eq.\ (\ref{integral}) would diverge for $g=0$).
This gap is
larger for the case of magnetic field along $C$-axis due to bare
easy-plane anisotropy in (\ref{modHam}).

\section{Comparison with experiment and discussion}

To compare our theory with an experiment we need to know the values of
the parameters of the Hamiltonian (\ref{modHam}) obtained from some independent
source. The inelastic neutron scattering experiments provide us with such a
source. As reported in \cite{Collins} $\tilde{J}=199$GHz and
$\tilde{D}=2.2$GHz.
These values were found as the best ones to fit the classical formulas for
spin-wave dispersion to the data of an inelastic neutron scattering. To get the
values of $J$ and $D$ in (\ref{modHam}) we have to take into account the
renormalizations by quantum fluctuations. Using (4) and (5) from
\cite{Chubukov}
we have: \[J=186 GHz,   D=1.3GHz \]
Using these values of $J$ and $D$ and the expressions (\ref{3par}) and
(\ref{3perp}) with $J'=0.22GHz$ from $H_c=\sqrt{48JJ'S^2}=3.9T$ we
have calculated the magnetization of \rmb\ for both $H \parallel C$ and $H
\perp
C$. The range of applicability of (\ref{3par}) and (\ref{3perp}) is
approximately
$H>4T$. For $H<4T$ the effects of three dimensional ordering and next order
corrections to the linear spin-wave theory have to be seriously taken into
account.

The magnetizations (\ref{3par}-\ref{3perp}) are plotted in Fig.1 together
with the experimental data for $M_{H\parallel C}$ and $M_{H\perp C}$. Also we
plotted the magnetizations calculated using classical formulas and the values
$J=199 GHz, D=2.2GHz$. We can see that the formulas (\ref{3par}) and
(\ref{3perp})
give both the much better absolute values of the magnetizations then the
classical
expressions and describe the anisotropy in magnetization. One can notice
also that
the experimental dependence $M_{H\parallel C}$ is a little bit nonlinear with
the slope increasing with H (this is not clearly seen from Fig.1 but can be
checked
by fitting the experimental data by linear functions in different ranges of
$H$).
This can be naturally explained by suppressing fluctuations with increasing $H$
and it is seen in (\ref{3par}).

The graphs obtained from (\ref{3par}-\ref{3perp}) in Fig.1 lie below
experimental
points. This probably can be corrected by the second order in $1/S$
corrections to
magnetization which will contribute with the sign opposite to the sign of the
first order corrections. The contribution of the higher order terms in $1/S$ is
not very small because the fluctuations are quite big in the experimental
range of parameters. For example $\frac{\Delta M_{H\perp C}}{M_{class}}\approx
30\%$ at $H=4.5T$.

\section{Conclusion}

The new effect of enhancement of anisotropy in quasi-one-dimensional spin
systems due to quantum and thermal fluctuations was found. It explains
the anisotropy in magnetization seen in experiments. The essence of the effect
is that the fluctuational part of magnetization is sufficiently big and is
determined mostly by logarithm of the gap in the spin wave spectrum. This gap
is anisotropic due to the bare easy-plane anisotropy. The logarithmical
dependence
of the fluctuational part of the magnetization on this gap (essentially the
divergence of one-dimensional fluctuations) leads to the strong "enhanced"
anisotropy of the magnetization. The same fluctuations explain a slightly
nonlinear
character of $M_{H\parallel C}$ dependence and the excessive values of
magnetization calculated  using classical formulas.

This theory can be applied to the compounds with easy plane anisotropy and
quasi-one-dimensional magnetic structure such as \rmb\ ,\cmb\ and \knc\
\footnote{For the magnetization data for \cmb\ taken from \cite{Jap,AbarBazh}
formulas (\ref{3par}-\ref{3perp}) give even the better
fit then for \rmb\ in Fig.1.}

\section{Acknowledgments}

The authors of this paper thank L.A. Prozorova, V.L. Pokrovskii, P. Wiegmann,
M. Zhitomirsky, M.F. Collins and especially A. Chubukov for fruitful
discussions.
We are indebted  to Yu.M. Tsipenyuk and M.F. Collins for sending us their
preprints before publication. One of us (O.P.) is grateful to A.N. Bazhan for
using his installation.

\newpage
\vspace{2cm}

\centerline{FIGURE CAPTIONS}

\vspace{1cm}
\noindent
{\bf Figure 1}  The magnetization $M_{long}$ parallel to the direction of
magnetic field as a function of field. Circles ($\circ$) denote
$M_{\parallel}(H)$ for $H\parallel C$  and squares ($\Box$) denote
$M_{\perp}(H)$ for $H\perp C$. Solid and dashed lines are the results of
calculation with parameters  $J=186$GHz, $D=1.3$GHz, $H_c=3.9$T, $J'=0.22$GHz
using formulas (\ref{3par} -\ref{3perp}) and classical theory \cite{Chubukov}
with $J=199$GHz, $D=2.2$GHz respectively.

\vspace{0.5cm}

\noindent
{\bf Figure 2}  The magnetization $M_{tr}$ perpendicular to the magnetic
field as
a function of $H$ with the field making small angles $\varphi$ with the basal
plane.  $\varphi =36^\circ$ (1), $\varphi =26^\circ$ (2), $\varphi =16^\circ$
(3), $\varphi =0.5^\circ$ (4), $\varphi =-14^\circ$ (5), $\varphi =-24^\circ$
(6), $\varphi =-34^\circ$ (7).

\vspace{0.5cm}
\noindent
{\bf Figure 3} The field dependence of the magnetization component $M_{tr}$ at
different temperatures with the field making angle $\varphi = 20^{\circ}$ with
the basal plane. T$=2.4$K (1), T$=4.2$K (2), T$=6.0$K (3), T$=7.1$K (4).

\vspace{0.5cm}

\noindent
{\bf Figure 4} The magnetization $M_{tr} (H=4.5$T, $\varphi =20^{\circ})$ is
shown  as a function of temperature. The dotted line is a guide to an eye.
\vspace{0.5cm}

\noindent
{\bf Figure 5}  Chosen quantization axes and coordinate systems.


\begin{thebibliography}{99}

\bibitem{Haldane}   Haldane F.D.M., Phys.Rev.Lett. {\bf 50} (1983), 1153.

\bibitem{Morra}     Morra R.M., Buyers W.J.L., Armstrong R.L. and Hirakawa
K.,
                    Phys.Rev.B. {\bf 38} (1988), 543.

                    Buyers W.J.L., Morra R.M., Armstrong R.L., Hogan M.J.,
                    Gerlach P. and Hirakawa K., Phys.Rev.Lett. {\bf 56} 1986,
                    371.

                    T.Ohyama and H.Shiba, Journ.Phys.Soc.Japan {\bf 62}
                    (1993), 3277 and reference there.

\bibitem{neutrCollins} B.D.Gaulin, M.F.Collins and W.J.L. Buyers, J.Appl.Phys.
                       {\bf 61} (1987), 3409.

B.D.Gaulin, T.E.Mason, M.F.Collins and J.Z.Larese, Phys.Rev.Lett.
             {\bf 62} (1989), 1380.
T.E.Mason, B.D.Gaulin and M.F.Collins, Phys.Rev.B {\bf 39} (1989), 586.

T.E.Mason, Y.S.Yang, M.F.Collins, B.D.Gaulin, K.N. Clausen and A.Harrison,
J.Magn.Mater. {\bf 104-107} (1992), 197.

\bibitem{cmbresonans}  Zaliznyak I.A., Prozorova L.A. and Petrov S.V.,
                       Sov.Phys.JETP. {\bf 70} (1990), 203.

\bibitem{Chubukov}  Chubukov A.V., Journ.Phys.C: Sol.St.Phys. {\bf 21}
                    (1988), 441.

\bibitem{Beba}      Kotyuzhanskii B.Ya. and Nikiforov D.V., J.Phys.:
                    Cond.Matter. {\bf 3} (1991), 385.

\bibitem{Jap}       Goto T., Inami T. and Ajiro Y., Journ.Phys.Soc.Japan
                    {\bf 59} (1990), 2328.

\bibitem{AbarBazh}  Abarzhi S.I., Bazhan A.N., Prozorova L.A., and
                    Zaliznyak I.A., J.Phys.: Cond.Matter. {\bf 4} (1992), 3307.

\bibitem{Glinka}    Glinka~C.J., Minkiewicz~V.J., Cox~D.E. and Khattak~C.P.,
                    Magnetism and Magnetic Materials, {\bf 18}th Ann.Conf.
                    (1973), 659.

\bibitem{Collins}   L.Heller, M.F.Collins, Y.S.Yang and B.Collier,
                    to be published in Phys.Rev.B

\bibitem{Kato_neutr}   T.Kato, T.Ishii, Y.Ajiro, T.Asano and S.Kawano,
                       Journ.Phys.Soc.Japan. {\bf 62} (1993), 3384.

\bibitem{Fink}      Von Fink H. and Seifert H.-J., Acta Cryst.
                    {\bf B38} (1982), 912.

\bibitem{Kato}      Kato~T., Yio~K., Hoshino~T., Mitsui~T. and Tanaka~H.,
                    Journ.Phys.Soc.Japan, {\bf 61} (1992), 275.

\bibitem{YMT}       J.B.Forsyth, R.M.Ibberson, Yu.M.Tsipenyuk and S.V.Petrov,
                    to be published at Phase Trans.

\bibitem{Kawano}    Kawano S., Ajiro Y. and Inami T., Journ. of Magnetism
                    and Magn.Mater. {\bf 104-107} (1992), 791.

\bibitem{my-stat}   A.N.Bazhan, I.A.Zaliznyak, D.V.Nikiforov, O.A.Petrenko,
                    S.V. Petrov and L.A.Prozorova. Sov.Phys.JETP. {\bf 76}
                    (1993), 342.

\bibitem{my-res}    I.M.Vitebskii, O.A.Petrenko, S.V.Petrov and L.A.Prozorova,
                    Sov.Phys.JETP. {\bf 76} (1993), 178.

\bibitem{Bazhan}    Bazhan A.N., Borovik-Romanov A.S. and Kreines N.M.,
                    Prib. Techn. Eksp. {\bf 1} (1973), 412.

\bibitem{Zaliznyak} I.A.Zaliznyak, Solid State Comm. {\bf 84} (1992), 573.

\bibitem{Welz} D.Welz, J.Phys.: Condens. Matter {\bf 5} (1993), 3643

\bibitem{Abanov}   A.Abanov, unpublished


\end{thebibliography}
\end{document}